\documentclass[prd,reprint,aps,tightenlines,preprintnumbers,amsmath,amssymb,showpacs,superscriptaddress]{revtex4-1}
\usepackage{amsmath,amssymb,graphicx,natbib}
\usepackage{graphicx}
\usepackage{dcolumn}
\usepackage{bm}
\usepackage{hyperref}
\usepackage{amsmath}

\def \beq{\begin{equation}}
\def \eeq{\end{equation}}

\def\lsim{\mathrel{\rlap{\lower4pt\hbox{\hskip1pt$\sim$}}
    \raise1pt\hbox{$<$}}}                
\def\gsim{\mathrel{\rlap{\lower4pt\hbox{\hskip1pt$\sim$}}
    \raise1pt\hbox{$>$}}}                
\newcommand{\bra}[1]{\langle #1|}
\newcommand{\ket}[1]{|#1 \rangle}

\begin{document}


\title{Muon Capture Constraints on Sterile Neutrino Properties}

\author{David McKeen}
\email{mckeen@uvic.ca}
\affiliation{Department of Physics and Astronomy, University of Victoria, Victoria, BC V8P 1A1, Canada}

\author{Maxim Pospelov}
\email{mpospelov@perimeterinstitute.ca}
\affiliation{Department of Physics and Astronomy, University of Victoria, Victoria, BC V8P 1A1, Canada}
\affiliation{Perimeter Institute for Theoretical Physics, Waterloo, ON N2J 2W9, Canada}

\date{\today}

\begin{abstract}
We show that ordinary and radiative muon capture impose stringent constraints on sterile neutrino properties.  In particular, we consider a sterile neutrino with a mass between $40$ to $80~{\rm MeV}$ that has a large mixing with the muon neutrino and decays predominantly into a photon and light neutrinos due to a large transition magnetic moment. Such a model  was suggested as a possible resolution to the puzzle presented by the results of the LSND, KARMEN, and MiniBooNE experiments \cite{Gninenko:2010pr}.  We find that the scenario with the radiative decay to massless neutrinos is ruled out by measurements of the radiative muon capture rates at TRIUMF in the relevant mass range by a factor of a few in the squared mixing angle. These constraints are complementary to those imposed by the process of electromagnetic upscattering and de-excitation of beam neutrinos inside the neutrino detectors induced by a large transition magnetic moment.  The latter provide stringent constraints on the size of the transitional magnetic moment between muon, electron neutrinos and $N$. We also show that further extension of the model with another massive neutrino in the final state of the radiative decay may be used to bypass the constraints derived in this work. 

\end{abstract}

\maketitle


\begin{section}{Introduction}

Muon capture has long served as a powerful probe of the weak interaction and of nuclear structure.  The closely related process of radiative muon capture (RMC), in which muon capture is accompanied by a photon emission, is greatly suppressed compared to ordinary muon capture (OMC), with a typical branching of 
$\Gamma_{\rm RMC}/\Gamma_{\rm OMC}\simeq 10^{-5}$.  The small rate of RMC allows for it to place tight constraints on models that would lead to increased rates of RMC. In this note we show a new application of the OMC and RMC results
as powerful constraints on sterile neutrino properties. We argue that in the kinematic range accessible 
to muon capture, it provides the best constraints on prompt radiative decays of sterile neutrinos.

Considerable experimental and theoretical effort has gone into the study of neutrino properties with 
medium energy proton beam experiments, such as LSND~\cite{Aguilar:2001ty}, KARMEN~\cite{Armbruster:2002mp}, and MiniBooNE~\cite{AguilarArevalo:2008rc,*AguilarArevalo:2010wv}. As it is well-known, 
no definitive conclusion can be made about the presence or absence of $\bar \nu_\mu \to \bar \nu_e$ 
neutrino oscillations, and claimed signals are difficult to reconcile with other measured properties 
of neutrino masses and mixings. While it appears not impossible that very small contributions to the 
fluxes of $\bar \nu_e$ might have been overlooked in the neutrino production simulations and while some backgrounds may not have been fully appreciated at MiniBooNE~\cite{Hill:2010zy}, it is 
nevertheless tempting to interpret these anomalies in terms of new physics. One such attempt was undertaken 
recently by Gninenko ~\cite{Gninenko:2009ks,Gninenko:2010pr}, who constructed a phenomenology-driven 
model of sterile neutrinos with sub-GeV mass, a large transitional magnetic moment, and a sizable 
mixing with the muon-type neutrino. The model is designed  as a possible explanation of some 
\cite{Gninenko:2009ks} or all ``unusual" neutrino signals \cite{Gninenko:2010pr}  at LSND and MiniBooNE.

Sterile neutrinos, among other very weakly interacting particles, can be a legitimate search 
object for different high-intensity/high-luminosity experiments (see, {\em e.g.} \cite{Bezrukov:2006cy}). 
Once produced, a heavy neutrino $N$ with $m_N < m_\pi$ would typically decay into three leptons,
and occasionally to $\nu\gamma$. Reference \cite{Gninenko:2010pr} considers a modification 
with a much enhanced transitional magnetic moment that considerably shortens the heavy neutrino
lifetime. To explain all the oscillation measurements while evading constraints from $K_{\mu 2}$ and $\pi_{\mu 2}$ decays Ref. \cite{Gninenko:2010pr} requires: 
(1) that the heavy neutrino, $N$, have a mass in the range $40~{\rm MeV}<m_N<80~{\rm MeV}$; (2) that its mixing with the muon neutrino have a strength $10^{-3}<\left|U_{\mu N}\right|^2<10^{-2}$; (3) that it decays primarily to light neutrinos and a photon; and (4) that its lifetime is $\tau_N<10^{-9}~{\rm s}^{-1}$.  Many different signatures in meson physics can be entertained with such an object and were described in Ref.~\cite{Gninenko:2010pr}. 
We notice that a sterile neutrino that mixes with the muon neutrino and decays dominantly to lighter neutrinos and a photon will greatly increase the RMC rate through the diagram in Fig.~\ref{fig:diagram}. Our calculations show that the whole mass and mixing range suggested by Gninenko is ruled out, if the final state of the radiative $N$ decay contains massless standard model (SM) neutrinos. On the other hand, we find that the model modified by 
introduction of another masive neutrino in the final state of the radiative decay of 
$N$ can pass the RMC constraints. 

In Sec.~\ref{sec:rmc}, we examine the constraints that RMC measurements have on sterile neutrinos that decay radiatively.  Section~\ref{sec:upscatt} contains estimates of the sterile neutrino electromagnetic production cross sections in a neutrino beam scattering on nuclei through a large transitional magnetic moment. We point out that such a process provides an additional source of constraints and should introduce important modifications to the analysis of Ref. \cite{Gninenko:2010pr}.   We discuss the effects that the production of sterile neutrinos through OMC on future $\mu\to e$ conversion experiments in Sec.~\ref{sec:mutoe}.  We conclude in Sec.~\ref{sec:conclusions}.

\begin{figure}
\begin{center}
\resizebox{60mm}{!}{\includegraphics{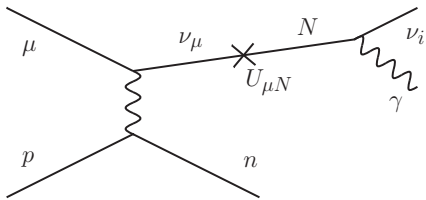}}
\caption{Diagram that leads to RMC induced by a radiatively decaying sterile neutrino $N$.  First, OMC produces 
an on-shell $N$ followed by the decay $N\to\gamma\nu$.}
\label{fig:diagram}
\end{center}
\end{figure}
\end{section}


\begin{section}{Radiative Muon Capture from a Sterile Neutrino}
\label{sec:rmc}
For simplicity we shall assume the same type of $V-A$ interaction 
governs the transition between the muon and the sterile neutrino. 
Then, we approximate the RMC rate induced by the production 
and subsequent radiative decay of a sterile neutrino, $N$, by the following formula:
\begin{align}
\frac{d\Gamma_{\rm RMC}}{dE_\gamma}&=\left|U_{\mu N}\right|^2\Gamma_{\rm OMC}\left(m_N\right){\cal B}\left(N\to \gamma \nu\right)f\left(E_\gamma\right)~,
\label{eq:rmcrate}
\end{align}
where $\Gamma_{\rm OMC}\left(m_N\right)$ is the rate of ordinary muon capture with a  massive sterile  
neutrino $N$ in the final state, modulo the mixing factor $U_{\mu N}$.  $f\left(E_\gamma\right)$ is the energy distribution of the photons,
\begin{align}
f\left(E_\gamma\right)=\frac{1}{\Delta E}-&\frac{2A_{LR}a}{\left(\Delta E\right)^2}\left(E_\gamma-E_{\rm av}\right)~~~,
\label{eq:spec}
\\
\Delta E=E_{\rm max}-E_{\rm min}~~,&~~E_{\rm av}=\frac{E_{\rm max}+E_{\rm min}}{2}~~~.
\nonumber
\end{align}
The parameter $a$ is defined by the angular distribution of the photon in the sterile neutrino's center-of-mass frame through
\begin{align}
\frac{1}{\Gamma\left(N\to \gamma \nu\right)}\frac{d\Gamma\left(N\to \gamma \nu\right)}{d\cos\theta}=\frac{1}{2}\left(1+a\cos\theta\right)~~~,
\label{eq:ang_dist}
\end{align}
where $\theta$ is the angle between the photon's direction and the spin of $N$.  In the most conservative case of a minimum of new fields beyond the SM, $N$ and $\nu$ are Majorana neutrinos, $a=0$, the decay of $N$ is isotropic and the energy distribution of photons in the lab frame is flat, {\it cf.} Eq.~(\ref{eq:spec}).  If the neutrinos are Dirac, the parameter $a$ may be in the range $-1$ to $1$, which introduces anisotropy in the rest frame of $N$ decay, and affects the energy distribution.  In the case of Dirac SM neutrinos with only left-handed couplings, $a\simeq1$~\cite{Li:1981um}.  $A_{LR}$ is the {\it helicity} asymmetry of the sterile neutrinos produced in OMC,
\begin{align}
A_{LR}=\frac{N_L-N_R}{N_L+N_R}
\label{eq:ALR}
\end{align}
with $N_{L,R}$ the number of left-(right-)handed sterile neutrinos produced.  Due to the assumed $V-A$ structure of the current producing the sterile neutrino, $A_{LR}=1$ for $m_N=0$, leading to no dilution of the photon anisotropy with respect to the sterile neutrino momentum.  However, $A_{LR}$ decreases as $m_N$ increases, partially washing out the anisotropy and leading to a flatter energy spectrum.  We show $A_{LR}$ as a function of $m_N$ and the photon energy spectrum for $a=1$ and $m_N=60,80~{\rm MeV}$ in Fig.~\ref{fig:spec_ALR} for OMC on hydrogen.

\begin{figure}
\begin{center}
\resizebox{80mm}{!}{\includegraphics{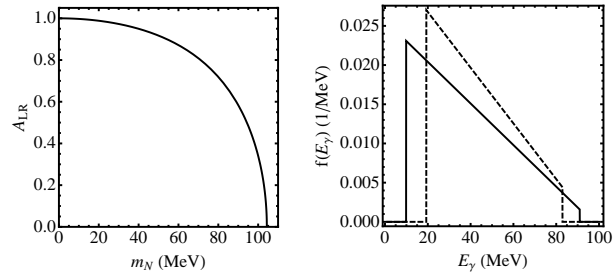}}
\caption{Left: the helicity asymmetry $A_{LR}$ as defined in Eq.~(\ref{eq:ALR}) as a function of $m_N$.  As expected, $A_{LR}=1$ for $m_N=0$ and $A_{LR}\to 0$ for $m_N\to m_\mu$.  Right: the photon energy spectrum [see Eq.~(\ref{eq:spec})] with $a=1$ [see Eq.~(\ref{eq:ang_dist})] for $m_N=60~{\rm MeV}$ (solid curve) and $m_N=80~{\rm MeV}$ (dashed curve).  The spectra do not go smoothly to zero near their respective maximum energies because $A_{LR}<1$ for both of these masses.  The plots are for $N$-producing OMC on hydrogen.}
\label{fig:spec_ALR}
\end{center}
\end{figure}

We focus on the case of muon capture on hydrogen since the theoretical uncertainties due to nuclear physics are lessened compared to capture on complicated nuclei.  Neglecting second class currents, the nuclear matrix element can be written as
\begin{align}
\label{eq:hadmatelem}
\bra{n}J_W^\alpha\ket{p}&=\bar{u}_n\left(p_2\right)\left[F_1\left(q^2\right)\gamma^\alpha+\frac{i}{2M_{np}}F_M\left(q^2\right)\sigma^{\alpha\beta}q_\beta\right.
\\
&\left.-g_A\left(q^2\right)\gamma^\alpha\gamma^5-\frac{1}{m_\mu}g_P\left(q^2\right)q^\alpha\gamma^5\right] u_p\left(p_1\right)~~~,
\nonumber
\end{align}
with $M_{np}=\left(m_n+m_p\right)/2$ and $q=p_2-p_1$.  It is necessary to calculate the rates for the cases where the muon and proton form a spin singlet and a spin triplet separately.  Using this matrix element, $\Gamma_{OMC}^{\rm sing}\left(m_N\right)$ and $\Gamma_{OMC}^{\rm trip}\left(m_N\right)$ can be found.  We use values of the form factors from Ref.~\cite{Bernard:1998gv,*Bernard:2000et} at $q^2=-0.88m_\mu^2+0.89m_N^2$ and assign an error of $\pm 20\%$ to $g_P$.  In Fig.~\ref{fig:RN}, we show $R_N^{\rm sing}$ and $R_N^{\rm trip}$ where each is the ratio of the OMC rate for a neutrino of mass $m_N$ to the OMC rate for a massless neutrino:
\begin{align}
R_N^{\rm sing}=\frac{\Gamma_{OMC}^{\rm sing}\left(m_\nu=m_N\right)}{\Gamma_{OMC}^{\rm sing}\left(m_\nu=0\right)}
\label{eq:rn}
\end{align}
and similarly for $R_N^{\rm trip}$.
\begin{figure}
\begin{center}
\resizebox{40mm}{!}{\includegraphics{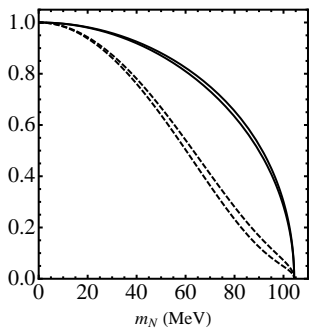}}
\caption{$R_N^{\rm sing}$ (solid curves) and $R_N^{\rm trip}$ (dashed curves) as defined in Eq.~(\ref{eq:rn}).  Each set of two curves is generated by varying $g_P$ by $\pm 20\%$.}
\label{fig:RN}
\end{center}
\end{figure}

E452 at TRIUMF measured the rate of RMC on hydrogen for photons with an energy $E_\gamma>E_\gamma^{\rm cut}=60~{\rm MeV}$ and found a RMC branching ratio~\cite{Wright:1998gi}
\begin{align}
R_\gamma=\left.\frac{\Gamma_{\rm RMC}}{\Gamma_{\rm tot}}\right|_{E_\gamma>60~{\rm MeV}}&=\left(2.10\pm 0.21\right)\times 10^{-8}~~~,
\label{eq:explimit}
\end{align}
where $\Gamma_{\rm tot}$ is the total muon decay rate.  Using this measurement, we can set a limit on $\left|U_{\mu N}\right|^2{\cal B}\left(N\to \gamma \nu\right)$ by requiring that the rate of RMC induced by the radiative decay of a sterile neutrino with $E_\gamma>60~{\rm MeV}$ not exceed this value, once we know the OMC rate for $N$ production.

In Ref.~\cite{Wright:1998gi} photons that originated outside of the hydrogen target were rejected.  Therefore, we must multiply our estimate of the RMC rate in Eq.~(\ref{eq:rmcrate}) by the probability, $P\left(m_N\right)$, that the decay $N\to\gamma\nu$ occurs in the target.  To estimate this probability, we calculate the average distance traversed by a sterile neutrino, $d_{\rm av}$, from a muon capture event to the edge of the target (a cylinder with a $16~{\rm cm}$ diameter and a $15~{\rm cm}$ height) assuming that the captures happen uniformly throughout the target and that the sterile neutrinos are emitted isotropically with respect to the detector orientation.  We find $d_{\rm av}=13.7~{\rm cm}$.  The probability is then estimated as
\begin{align}
P\left(m_N\right)&=1-\exp\left(\frac{-d_{\rm av}}{\gamma v \tau}\right)~~~.
\label{eq:prob}
\end{align}
This probability is approximately 18\% for $m_N = 40~{\rm MeV}$ and 43\% for $80~{\rm MeV}$ if $\tau\left(N\to\gamma\nu\right)=10^{-9}~{\rm s}$.

To accurately compare with experiment, the relative fractions of different $\mu p$ bound states in the target used in Ref.~\cite{Wright:1998gi} must be taken into account.  Muons are captured in the atomic singlet and triplet states.  In the target used, liquid hydrogen at 16 K, the atomic triplets quickly transition to the atomic singlet.  These atomic $\mu p$ singlets tend to form molecular $p \mu p$ states that can either be in the ortho (proton spins parallel) or para states (proton spins antiparallel).  This results in the muons being captured from a mixture of atomic singlet and molecular ortho and para states.  Reference~\cite{Wright:1998gi} found the relative occupancies to be
\begin{align}
f_{\rm sing}=0.061~,~~~f_{\rm para}=0.085~,~~~f_{\rm ortho}=0.854~.
\end{align}
The molecular OMC rates can be related to the atomic ones,
\begin{align}
\Gamma_{\rm OMC}^{\rm para}&=2\gamma_P\frac{1}{4}\left(\Gamma_{\rm OMC}^{\rm sing}+3\Gamma_{\rm OMC}^{\rm trip}\right)~~~,
\\
\Gamma_{\rm OMC}^{\rm ortho}&\simeq2\gamma_O\frac{1}{4}\left(3\Gamma_{\rm OMC}^{\rm sing}+\Gamma_{\rm OMC}^{\rm trip}\right)~~~,
\end{align}
where the approximately equal sign in the ortho rate reflects the fact that the ortho molecular state is thought to be dominantly spin-1/2 with little spin-3/2 contribution~\cite{Bakalov:1979jj}.  The wavefunction corrections are $\gamma_P=0.5733$ and $\gamma_O=0.500$~\cite{Wessel:1964zz,*Bakalov:1980fm}.  The total OMC rate is then
\begin{align}
\Gamma_{\rm OMC}=f_{\rm sing}\Gamma_{\rm OMC}^{\rm sing}+f_{\rm para}\Gamma_{\rm OMC}^{\rm para}+f_{\rm ortho}\Gamma_{\rm OMC}^{\rm ortho}~~~.
\end{align}
Knowing the atomic singlet and triplet OMC rates as functions of $m_N$ and the relative abundances of different 
muon-containing modifications of hydgrogen, we calculate $\Gamma_{\rm OMC}\left(m_N\right)$.  This rate, the relationship in Eq.~(\ref{eq:rmcrate}), the measurement in Eq.~(\ref{eq:explimit}), the energy spectrum in Eq.~(\ref{eq:spec}), the probability in Eq.~(\ref{eq:prob}), and the assumption of a lifetime $\tau\left(N\to\gamma\nu\right)<10^{-9}~{\rm s}$ allow us to limit $\left|U_{\mu N}\right|^2{\cal B}\left(N\to \gamma \nu\right)$.  This limit is shown in Fig.~\ref{fig:ulimit} for $a=-1,0,1$.  For a branching ratio ${\cal B}\left(N\to \gamma \nu\right)\simeq 1$, this implies that $\left|U_{\mu N}\right|^2\lsim 7.9\times10^{-4}$ over the range $40$ to $80~{\rm MeV}$ for $a=1$, slightly below the mixing required in Ref.~\cite{Gninenko:2010pr}.  For $a=0,-1$, with ${\cal B}\left(N\to \gamma \nu\right)\simeq 1$, the limit over this mass range is stronger, $\left|U_{\mu N}\right|^2\lsim \left(3.3,2.1\right)\times10^{-4}$ respectively.  We note that this is a conservative upper limit since we have simply required that the number of RMC events due to the radiative decay of the sterile neutrino $N$ not exceed the total number of events seen. In practice, $N$-induced RMC occurs along with
standard RMC, and requiring that the sum of the two contributions agrees with experimentally observed rates 
would additionally strengthen the bound by a factor of $\sim 2-3$.  We also notice that our estimate of the OMC rate is on the slightly low side ($\Gamma_{\rm OMC}^{\rm sing}=687-711~{\rm s}^{-1}$ as $m_N\to 0$ with our range of nuclear form factors as compared to the experimental value of $\Gamma_{\rm OMC}^{\rm sing}=725.0\pm17.4~{\rm s}^{-1}$~\cite{Andreev:2007wg}) which also makes our limit on 
$\left|U_{\mu N}\right|^2$ more conservative.

Our calculations also show how the RMC limits can be weakened or 
avoided at the price of slight modification of the orginal 
model in  Ref.~\cite{Gninenko:2010pr}. 
In order to do that, one can further ``downgrade" the photon spectrum by assuming a massive neutrino 
$N^\prime$ in the final state of the radiative decay, $N\to N^\prime \gamma$. 
The photon energy is then decreased by the factor $\left(1-m_{N^\prime}^2/m_N^2\right)$.  If the $N^\prime$ is massive enough, the maximum photon energy could be below the experimental cut of $60~{\rm MeV}$.  For $m_N=60~{\rm MeV}$, this requires $m_{N^\prime}\geq 35~{\rm MeV}$.

One should also be aware, of course, of the 
discrepancy in the RMC-extracted value for the $g_P$ form factor compared to the chiral perturbation theory 
calculations. The experimental value for $g_P$ from the standard RMC seems to be in excess of the standard model (SM) prediction
at a $\sim 30\%$ level~\cite{Bernard:2001rs}. We note in passing that one could speculate that 
such discrepancy could also originate from the radiatively decaying 
sterile neutrino, with a squared mixing angle at the level of ${\cal O}(10^{-4})$. 
\begin{figure}
\begin{center}
\resizebox{60mm}{!}{\includegraphics{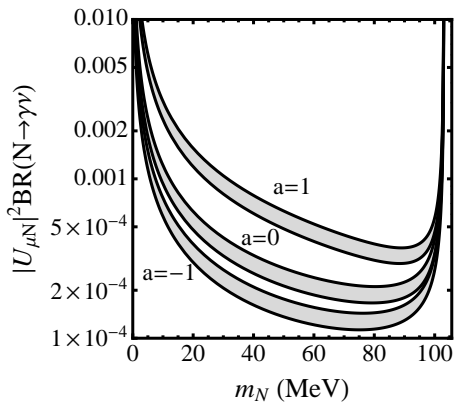}}
\caption{The range of limits on $\left|U_{\mu N}\right|^2{\cal B}\left(N\to \gamma \nu\right)$ with $m_\nu=0$ for $a=-1,0,1$ (from bottom to top) implied by the RMC rate measured in Ref.~\cite{Wright:1998gi} for a sterile neutrino of mass $m_N$ with lifetime $\tau\left(N\to\gamma\nu\right)<10^{-9}~{\rm s}$.  The ranges are generated by varying $g_P$ by $\pm 20\%$ and the measurement from Ref.~\cite{Wright:1998gi} by $\pm 1\sigma$.  The $a=1$ case is least constrained since photons are then preferentially emitted opposite the direction of the sterile neutrino's momentum, {\it cf.} Eq.~(\ref{eq:spec}).  The limits weaken for smaller $m_N$ since the probability of the radiative decay occurring within the target in Eq.~(\ref{eq:prob}) decreases.}
\label{fig:ulimit}
\end{center}
\end{figure}
\end{section}

\begin{section}{Producing $N$ through the Magnetic Moment}
\label{sec:upscatt}
The radiative decay rate of the sterile neutrino to SM neutrinos is given by
\begin{align}
\Gamma\left(N\to\gamma\nu\right)=\frac{\pi\alpha}{8}\sum_i\left(\frac{\mu_{\rm tr}^{i}}{\mu_B}\right)^2\frac{m_N^3}{m_e^2}~~~,
\end{align}
where $\mu_{\rm tr}^{i}$ is the transition magnetic moment between $N$ and the  $i^{\rm th}$ lighter neutrino.  Factors of $m_e$ in the above formula are due to a conventional normalization of magnetic moments 
on the Bohr magneton $\mu_B$. If one identifies $\nu$ with the three light neutrinos of the SM and assumes a common value of the transition magnetic moment between $N$ and each of them, a radiative decay lifetime of $10^{-9}~{\rm s}$ requires 
\begin{align}
\mu_{\rm tr}^{e,\mu,\tau}=9.6\times10^{-9}\mu_B~~~.
\label{eq:magmom}
\end{align}
A transition magnetic moment this large raises the possibility that these sterile neutrinos could be produced by neutrino beams electromagnetically upscattering on nuclei.  A similar production mechanism was considered in \cite{Gninenko:1998nn}.  Muon or electron (anti)neutrinos scattering electromagnetically on a target of spin-1/2 with charge $Ze$ will produce sterile neutrinos through the process $\nu_{e,\mu}Z\to NZ$ with the cross section (ignoring subdominant scattering on the nuclear magnetic dipole moment)
\begin{align}
\frac{d\sigma}{dT}&=\frac{4\pi\alpha^2}{m_e^2}\left(\frac{\mu_{\rm tr}^{e,\mu}}{\mu_B}\right)^2 Z^2 \left|F\left(q^2\right)\right|^2\Bigg\{\frac{1}{T}-\frac{1}{E_\nu}
\\
&+\frac{m_N^2}{4 M E_\nu^2}\left(1-\frac{M+2E_\nu}{T}\right)+\frac{m_N^4}{8 M^2 E_\nu^2}\left(\frac{1}{T}-\frac{M}{T^2}\right)\Bigg\},
\nonumber
\end{align}
where $M$ is the mass of the target, $T$ is its recoil kinetic energy and $F\left(q^2\right)$ its electric form factor, and $E_\nu$ is the energy of the light neutrino.  For a scalar target ({\em e.g.} $^{12}$C nucleus) this cross section becomes
\begin{align}
\frac{d\sigma}{dT}&=\frac{4\pi\alpha^2}{m_e^2}\left(\frac{\mu_{\rm tr}^{e,\mu}}{\mu_B}\right)^2 Z^2 \left|F\left(q^2\right)\right|^2\Bigg\{\frac{1}{T}-\frac{1}{E_\nu}+\frac{T}{4E_\nu^2}
\nonumber
\\
&+\frac{m_N^2}{4 M E_\nu^2}\left(\frac{1}{2}-\frac{M+2E_\nu}{T}\right)-\frac{m_N^4}{8 M E_\nu^2}\frac{1}{T^2}\Bigg\}~~~.
\end{align}
These expressions agree with those found in Ref.~\cite{Vogel:1989iv} in the limit that $m_N\to0$.  The energy of the sterile neutrino is $E_N=E_\nu-T$.  At LSND, this process would lead to a deposit of electromagnetic energy through $\nu~^{12}C\to N~^{12}C \to\gamma\nu~^{12}C$.  Using a charge radius for $^{12}C$ of $2.46~{\rm fm}$~\cite{Reuter:1982zz}, we show $\sigma\left(\nu~^{12}C\to N~^{12}C\right)$ as a function of the incoming neutrino energy for $m_N=60~{\rm MeV}$ and $\tau\left(N\to\gamma\nu\right)=10^{-9},~10^{-10},~10^{-11}~{\rm s}$ in Fig.~\ref{fig:crosssec}, assuming a common transition moment between $N$ and the light neutrinos.  Using the flux of $\nu_\mu$ and $\nu_e$ from pions decaying in flight in Ref.~\cite{Aguilar:2001ty}, one can estimate the number of sterile neutrinos produced in the LSND target through this process,
\begin{align}
N\left(N\right)\simeq &4.9\times10^5 \left(\frac{\mu_{\rm tr}^\mu}{10^{-8}\mu_B}\right)^2+330 \left(\frac{\mu_{\rm tr}^e}{10^{-8}\mu_B}\right)^2,
\end{align}
which would mimic $\nu_e C\to e^- N$ ($N$ refers to nitrogen here) events after the radiative decay of the sterile neutrino.  The fitted number of events of this type was 18.0, indicating that a radiative decay lifetime for $N$ of $10^{-9}~{\rm s}$ or shorter is ruled out if there is a common transition magnetic moment to the three light neutrinos.

The production mode due to the transition magnetic moment and strong 
constraints on the overall number of electron-like events in LSND may be avoided if
transitional magnetic moments are flavor-nonuniversal.  One could avoid both the problem of exessive $N$ production in the neutrino detector and strong constraints on 
$\left| U_{\mu N}\right|^2$ in Sec.~\ref{sec:rmc} using a hierarchy of magnetic moments,
\begin{align}
\mu_{\rm tr}^{\tau} \gg \mu_{\rm tr}^\mu,~\mu_{\rm tr}^e~~~,
\end{align}
and still possibly provide an explanation to at least some of the experimental anomalies.  
If $\mu_{\rm tr}^\mu$ is decreased by two orders of magnitude from the value in Eq.~(\ref{eq:magmom}) and $\mu_{\rm tr}^e$ by a factor of at least several so that $\Gamma\left(N\to\gamma\nu_\mu\right)\lsim 10^{5}~{\rm s^{-1}}$ and $\Gamma\left(N\to\gamma\nu_e\right)\lsim 10^{7}~{\rm s^{-1}}$, the 
total rate of $\nu_e C\to e^- N$ events could be made consistent with the LSND bound.  
If a large transition magnetic moment between $N$ and $\nu_\tau$ is maintained ($\mu_{\rm tr}^\tau\gsim 10^{-8} \mu_B$), 
\begin{align}
10^{-11}~{\rm s}\lsim\tau\left(N\to\gamma\nu_{\tau}\right)\lsim 10^{-9}~{\rm s}~~~,
\end{align}
the sterile neutrinos would then be produced through their transition magnetic moment with $\nu_e$ or $\nu_\mu$ while they would decay promptly to $\gamma\nu_\tau$.  The mixing $\left| U_{\mu N}\right|^2$ could then be reduced to agree  with the limit from RMC.  While it appears possible to explain the strength of the MiniBoone signal that way, 
a detailed study of the LSND signal is a lot more complicated. It would involve some nuclear modeling to determine whether the $\nu_\mu \to N$ up-scattering process could additionally dislodge a neutron from the nucleus with large enough probability to look like the LSND signal without disturbing other measurements.  It is also not immediately clear without further investigation if the angular distribution of photons from this production mechanism would be compatible with the signal. Finally, one should also assess the constraints on the transitional magantic moments of 
atmsopheric $\tau$ neutrinos imposed by the super-Kamiokande electron-like rates.  This model's motivation is purely phenomenological and the question of making a sensible UV completion of it must also be addressed. 

\begin{figure}
\begin{center}
\resizebox{60mm}{!}{\includegraphics{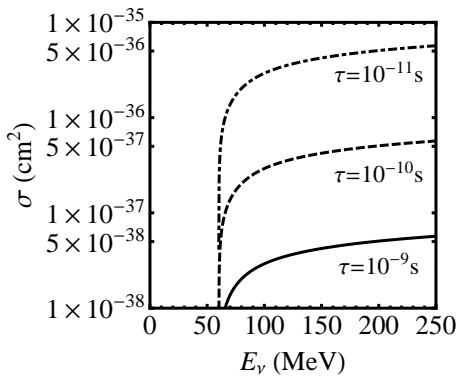}}
\caption{$\sigma\left(\nu~^{12}C\to N~^{12}C\right)$ for $m_N=60~{\rm MeV}$ and $\tau\left(N\to\gamma\nu\right)=10^{-9},~10^{-10},~10^{-11}~{\rm s}$ as functions of the incoming neutrino energy.  We have assumed a common transition magnetic moment between $N$ and the light neutrinos, $\mu_{\rm tr}/\mu_B=9.6\times10^{-9},~3.0\times10^{-8},~9.6\times10^{-8}$ respectively.}
\label{fig:crosssec}
\end{center}
\end{figure}
\end{section} 

\begin{section}{Sensitivity in $\mu\to e$ Conversion}
\label{sec:mutoe}
Searches for sterile neutrinos with a mass less than that of the muon that mix with muon neutrinos are particularly suited to experiments looking for $\mu \to e$ conversion in the presence of a nucleus due to the large number of muon captures. Large $Z$ nuclei are used as targets, greatly increasing the muon capture rate.  In aluminum, for example, the capture rate and the free decay rate are roughly equal, and for larger $Z$ the capture rate begins to dominate.

The proposal for the Mu2e experiment at FNAL~\cite{Carey:2008zz} envisions around $10^{17}$ muon caputures per year, producing roughly
\begin{align}
\sqrt{1-\frac{m_N^2}{m_\mu^2}}\left|U_{\mu N}\right|^2\times10^{17}
\end{align}
sterile neutrinos.  Using a $^{48}{\rm Ti}$ target, branching ratios of the order $10^{-16}$ would be probed.  Mixing angles
\begin{align}
\left|U_{\mu N}\right|^2\gsim\frac{10^{-16}}{\epsilon\sqrt{1-m_N^2/m_\mu^2}}
\end{align}
could be observed, with $\epsilon<1$ describing the fraction of decays that would pass experimental cuts, nuclear physics, and efficiencies that depend on the details of the experimental setup and model under consideration, 
{\em e.g.} the lifetime of $N$.  For small $U_{\mu N}$, of course, standard RMC becomes dominant and information about the spectral shape is necessary to increase sensitivity.  The nuclear physics in high-$Z$ materials is also less well known than in the case of hydrogen considered in Sec.~\ref{sec:rmc}. Another benefit from such experiments is that sterile neutrinos with less exotic decay modes, like $\nu e^+e^-$, could also be explored.
\end{section}

\begin{section}{Conclusions}
\label{sec:conclusions}
In this note, we have shown that RMC can stringently constrain models of new physics. It is particularly suited 
for putting constraints on models of sterile neutrinos in which they promptly decay with a photon and the massless 
neutrino in the final state.  The large transition magnetic moment required for a sterile neutrino to decay within a typical detector volume means that it can also be an efficient production mechanism.  This greatly constrains the transition moment with electron and especially muon neutrinos.  Further study is required to investigate whether this production mechanism could provide exotic but viable explanations of the neutrino experiment anomalies.  
As for the model that partially motiviated this study, Ref. \cite{Gninenko:2010pr}, we find that the part of 
the parameter space with $m_N$ in the interval between 40 and 80 MeV, and the square of mixing angle 
above $10^{-3}$ is in conflict with the RMC constraints. It is possible, however, to modify the model
and avoid contraints from RMC and magnetic moment induced production by introducing 
more degrees of freedom beyond SM. We find that a radiative decay into another massive state, 
$N\to N^\prime \gamma$ may provide a remedy. 

{\em Note added}: In a recent comment on our paper \cite{Gninenko:2010nv}, 
Gninenko finds numerically different results from our 
constraints. Our current version includes the effects of possible angular anisotropies in the 
decay of $N$. We believe that the remaining differences between our and Gninenko's results might be
attributed to what appears to be only one helicity of $N$ in the final state of OMC in Ref.~\cite{Gninenko:2010nv}, while both should be present.

\end{section}

\bibliography{ref}

\end{document}